\documentclass[12pt]{article}
\usepackage{amsmath, amsfonts,euscript,bbm}
\usepackage{epsfig, cite}
\usepackage{color}
\usepackage{ifthen}
\usepackage{graphicx}
\usepackage{setspace} 
\newboolean{Notes}\setboolean{Notes}{true}
\newcommand{\notes}[1]{\ifthenelse{\boolean{Notes}}{\textcolor{black}{#1}}{}}

\newcommand{\skyp}[1]{}

\begin{document}

\hskip 4in
\vbox{\baselineskip12pt \hbox{FERMILAB-PUB-07-063-A}}
\bigskip\bigskip\bigskip\bigskip

\centerline{\Large A New Spin on Quantum Gravity}
\bigskip
\bigskip
\bigskip
\centerline{\bf Mark G. Jackson}
\medskip
\centerline{Particle Astrophysics Center}
\centerline{Fermi National Accelerator Laboratory}
\centerline{Batavia, Illinois 60510}
\centerline{\it markj@fnal.gov}
\bigskip
\centerline{\bf Craig J. Hogan}
\medskip
\centerline{Departments of Physics and Astronomy}
\centerline{University of Washington, Seattle}
\centerline{Seattle, Washington 98195}
\centerline{\it hogan@u.washington.edu}
\bigskip
\bigskip
\begin{abstract}
We suggest that the (small but nonvanishing) cosmological constant, and the holographic properties of gravitational entropy,  may both reflect unconventional quantum spin-statistics at a fundamental level.  This conjecture is motivated by the nonlocality of quantum gravity  and the fact that spin is an inherent property of spacetime.  As an illustration we consider the `quon' model which interpolates between fermi and bose statistics, and show that this can naturally lead to an arbitrarily small cosmological constant. In addition to laboratory tests, we briefly discuss the possible observable imprint on cosmological fluctuations from inflation.\end{abstract}
\bigskip
\bigskip
\bigskip
\bigskip
\bigskip
\bigskip
\centerline{{\em This essay received an Honorable Mention in the}}
\centerline{{\em 2007 Gravity Research Foundation Essay Competition}}
\newpage            
\baselineskip=18pt
\doublespacing
Two of the biggest puzzles in theoretical physics are related to gravity:
\begin{enumerate}
\item The (nearly) vanishing of the cosmological constant \cite{Weinberg:1988cp} \cite{Riess:1998cb}.  A precise vanishing arises naturally in supersymmetry, whereby the vacuum energy of each boson is cancelled by that of its fermionic superpartner, and vice-versa.  We do not observe these superpartners, and the cosmological constant does not precisely vanish; possibly both effects are associated with the breaking of a fundamental supersymmetry. 

\item Why inclusion of gravity in a quantum system produces far fewer degrees of freedom than naively predicted with no gravity, in effect reducing the dimensionality of a quantum theory by one (and thus earning the name `holography') \cite{'tHooft:1993gx} \cite{Susskind:1994vu}.  In an extreme case,  when a system has too much localized energy it   becomes a black hole, whose entropy is proportional not to its 3D volume but to its 2D area.  Such dimensional-reducing behavior calls for some new physical principle outside the canon of usual local quantum field theory.
\end{enumerate}
It has been suggested \cite{Fischler:1998st} \cite{Bousso:1999xy} \cite{Cohen:1998zx} that there is a close connection between these two puzzles.  Here we propose a new connection between them, based on a conjectured violation of the usual Pauli spin-statistics, where integral-spin particles are in symmetric wavefunctions and half-integral spin particles are in antisymmetric wavefunctions.  Such a theory is not local nor Lorentz-invariant, but since we expect neither in a quantum theory of gravity, we find this to be acceptable in some as-yet-untested physical regime.

This spin-statistics violation is implemented as follows.  An idea previously suggested by one of us \cite{Jackson:2005ue} was that gravity can be modeled by allowing the $[x,p] = i \hbar$ commutator to be a function of energy scale, so that the effective unit of quanta changes.  Here we employ a similar but more concrete commutator modification, whereby gravity modifies the relation $a_i a_j^\dagger \pm a_j^\dagger a_i = \delta_{ij}$ for bosons (-) and fermions (+).  Various examples of generalizations to usual fermi/bose commutators have been studied in the literature, but we will specifically consider the `quon'\footnote{rhymes with `muon'} model developed by Greenberg \cite{Greenberg:1989ty} \cite{Greenberg:1991ec},
\[ a_i a_j^\dagger - q a_j^\dagger a_i = \delta_{ij}. \]
Bose and fermi statistics are recovered in the limits of $q=+1$ and $q=-1$, respectively, but $-1 \leq q \leq 1$ are all valid theories.  

We propose that in some `holographic limit' $q$ approaches very close to zero for \emph{all} particles, regardless of spin.  As emphasized by Greenberg the $q=0$ algebra is not only technically simple, it is in some sense the most fundamental: all theories $-1 \leq q \leq 1$ can be constructed out of the $q=0$ system.  It also possesses a notable property which is apparently unknown in the literature: the vacuum energy vanishes identically, for any system, for any number of fields!  In supersymmetry (for which the bosonic and fermionic states both have identical energy $\epsilon_i$) the bosonic modes have total energy $E_i = \epsilon_i (a_i^\dagger a_i + \frac{1}{2}) $ while fermionic modes have total energy $E_i =  \epsilon_i (a_i^\dagger a_i - \frac{1}{2})$.  The vacuum energies then cancel identically,  $\Lambda_i =  \epsilon_i ( \frac{1}{2} - \frac{1}{2} ) = 0$.  In the case of $q=0$, however, there is simply no vacuum energy term present:
\[ E_i = \epsilon_i ( a_i^\dagger a_i + \sum_k a_k^\dagger a_i^\dagger a_i a_k + \sum_{k,l} a_l^\dagger a_k^\dagger a_i^\dagger a_i a_k a_l + \ldots ) . \]
If all fields were to behave as though $q\sim0$ then the cosmological constant would be arbitrarily small.

Why should such a spin-statistics violation happen?  It seems natural to expect such a breakdown from quantum gravity.  Spin is the way in which a particle transforms under spacetime rotations, and quantum gravity represents fluctuations in spacetime.  Thus it is reasonable to expect that the effective $q$ may change, much as a coupling constant `flows' due to quantum loop corrections.  As $q$ deviates from $\pm 1$ the physics becomes nonlocal although it still possesses many properties (such as CPT, clustering, Wick's theorem, etc.) which appear to make it a sensible quantum field theory.  Since $q=0$ is `maximally nonlocal' one might expect this is the one favored by holography.

Unlike many ideas in quantum gravity, quonic behavior causes observable effects in the laboratory via violations in spin-statistics. There are existing bounds on the violation of Fermi statistics, first performed by Ramberg and Snow \cite{Ramberg:1988iu} and more recently by the VIP collaboration \cite{VIP}:
 \[ \frac{1+q_F}{2} \leq 4.5 \times 10^{-28}. \]
There are also plans for an  experiment which will measure the time-variation of such violations \cite{neon}.  

However, since the energy scale of our conjectured violations probably lies well outside the reach of laboratory experiments, the  best test of the quon-flow hypothesis might come from  cosmological structure, now measured with considerable detail and precision in   maps of the cosmic background anisotropy.  The largest scale  pattern on the sky records  a direct fossil imprint on the metric  by quantum fluctuations in  bosonic quantum fields (either the spin-zero inflaton or the spin-2 graviton) at the epoch of cosmic inflation, well beyond the energy and length scales of laboratory tests, and  close enough to the Planck time that quantum gravity effects may be noticeable. In particular one of us has demonstrated \cite{Hogan:2003mq} that the Hilbert space of  standard field-theory modes is far too large to be consistent with holographic bounds.   Although standard fields can give the correct averaged power spectrum,  for consistency the relic classical observables must include new correlations reflecting the nonlocal character of the true quantum behavior imposed by holography when spacetime degrees of freedom are also included.  As we have seen, quons define a  nonlocal quantum system that nevertheless has field theory as a sensible limiting case.

In addition, the near-vanishing of the quonic cosmological constant suggests that for $|q| \ll 1$, fermionic and bosonic degrees of freedom are intertwined such that even the mean square zero-point fluctuations of the vacuum nearly cancel on average; these  are the same fluctuations that when ``frozen out'' give the cosmic perturbations, so  quonic fundamental modes with $q$ close to zero might naturally explain the smallness of cosmological perturbations (a dimensionless quantity, sometimes aptly named $Q$,  measured to be around $10^{-5}$).   There has not up to now been a convincing way to extend field theory inflationary calculations to accommodate the holographic bounds; a new theory of  ``quonic inflation,'' based on quantized quon fields in an expanding classical background, would allow quantitative estimates of new phase correlations  in the patterns of sky maps, and possibly even a derivation of $Q$ from first principles. 

Although the connections of quons to holography and inflation appear new, we are not the first to suggest that modified spin statistics may be needed to explain the mysteries of vacuum energy. Indeed, the greatest master of spin, Wolfgang  Pauli himself, put it even more strongly, commenting that the infinite zero point energy of the vacuum derived from the quantized field ``is an indication that a fundamental change in the concepts underlying the present theory of quantized fields will be necessary.'' \cite{paulinobel}

We would like to thank C. Petrascu, E. Ramberg and J. Santiago for useful discussions and comments.  The work of MGJ was supported by the DOE and the NASA grant NAG 5-10842 at Fermilab.  CH would like to acknowledge the hospitality of Fermilab and the Enrico Fermi Institute at the University of Chicago.

\end{document}